\documentstyle[epsfig,referee]{mn2e}
\title[Electron spectral indices in GRBs \& other relativistic sources]
{No Universality for Electron's Power-Law Index (p) in
Gamma-Ray Bursts and Other Relativistic Sources}

\author[Shen, Kumar \& Robinson]
{Rongfeng Shen \thanks{E-mail: rfshen@astro.as.utexas.edu.}
, Pawan Kumar, and Edward L. Robinson\\
Department of Astronomy, University of Texas at Austin,
1 University Station C1400, Austin, TX 78712, USA}

\date{}

\begin{document}

\maketitle

\begin{abstract}

The Gamma-Ray Burst (GRB) prompt emission is believed to be from
highly relativistic electrons accelerated in relativistic shocks.
From the GRB high-energy power-law spectral indices $\beta$ observed
by the Burst and Transient Source Experiment (BATSE) Large Area
Detectors (LAD), we determine the spectral index, $p$, of the
electrons' energy distribution. Both the theoretical calculations
and numerical simulations of the particle acceleration in
relativistic shocks show that $p$ has a universal value $\approx
2.2-2.3$. We show that the observed distribution of $p$ during GRBs
is not consistent with a $\delta$-function distribution or an
universal $p$ value, with the width of the distribution $\ge$ 0.54.
The distributions of $p$ during X-ray afterglows are also
investigated and found to be inconsistent with a $\delta$-function
distribution. The $p$-distributions in blazars and pulsar wind
nebulae are also broad, inconsistent with a $\delta$-function
distribution.

\end{abstract}


\section{Introduction}

GRBs are observed to have non-thermal spectra during its prompt
emission phase (Band et al. 1993). It is widely believed that the
synchrotron radiation and/or the inverse Compton scattering are
the likely emission mechanism(s) for GRB's prompt hard X-ray and
$\gamma$ ray emission. The electrons accounting for these
emissions are thought to be accelerated in relativistic shocks in
GRBs. According to the shock diffusive acceleration model,
particles are accelerated when they repeatedly cross a shock
front, and the competition between the particle's energy again and
escape probability per shock crossing cycle leads to a power-law
spectrum for the particles:
\begin{equation} N(\gamma) d\gamma \propto \gamma^{-p} d\gamma\,\,,  \end{equation}
where $\gamma$ is the Lorentz factor of the particle 
(e.g., Blandford \& Ostriker 1978).
For non-relativistic shocks, the value of $p$ depends on the the compression
ratio of the flow stream across the shock; while in relativistic or
ultra-relativistic shocks, which is most likely the case in GRBs, analytical and
numerical studies show that $p$ has an  ``universal'' value, $\approx 2.2-2.3$
(Kirk et al. 2000, Achterberg et al. 2001, Bednarz \& Ostrowski 1998,
Lemonine \& Pelletier 2003).

The work reported in this paper is to investigate the ``universality''
of the power-law index $p$ for GRBs, which we calculate directly
from the high-energy ($0.1 - 2$ MeV) photon spectrum of GRBs
(Preece et al. 1998, 2000), assuming the spectrum is from
synchrotron or synchrotron self-inverse-Compton emission of
the power-law distributed highly
relativistic electrons, using the relations between $p$
and the spectral index , $\beta$, of the high-energy power-law photon
spectrum.

In \S2, we describe the GRB spectral data set used and the
process of determining the parent $p$-distribution.
In \S3, we examine the contributions from the
spectral fit procedure and the time averaging effect to the dispersion of the
parent distribution of $p$.
The $p$-distributions derived from BeppoSAX GRBs and from
HETE-2 GRBs, X-ray flashes and X-ray rich GRBs are presented in \S4.
We determine the $p$-distributions
for X-ray afterglows in \S5 and for blazars and pulsar wind nebulae in \S6.
A summary and discussions are given in \S7.


\section{The distribution of $p$ in GRBs}

\subsection{The GRB spectral sample}

For our analysis, we use the BASTE GRB Spectral Catalog presented by
Preece et al. (2000). In the catalog, the time sequences of spectral
fit parameters for 156 bright bursts are presented, using mostly the
high energy and time resolution data from the Large Area Detectors
(LAD), which covers an energy range of typically 28 - 1800 keV. All
bursts have at least 8 spectra in excess of 45 $\sigma$ above
background. The spectral models used in fit are (i) `Band'
function; (ii) Comptonized spectral model (a power-law with an
exponential cut-off); (iii) Broken Power-Law model; and (iv)
Smoothly Broken Power-Law model. The `Band' function, the one used
most frequently, is an empirical function (Band et al. 1993)
\begin{displaymath}
N(E) = A \left\{\begin{array}{ll} (E/100)^{\alpha} \exp[-E(2+\alpha)/E_{peak}], & E < \frac{\alpha-\beta}{2+\alpha}E_{peak} \\
                    \Big [\frac{(\alpha-\beta)E_{peak}}{100(2+\alpha)}\Big ]^{\alpha-\beta} \exp(\beta-\alpha) (E/100)^{\beta}, &
                    E \ge \frac{\alpha-\beta}{2+\alpha}E_{peak}
\end{array}\right.
\end{displaymath}
where $N(E)$ is the photon counts, $A$ is the amplitude, $\alpha$ is
the low-energy spectral index, $\beta$ the high-energy spectral
index, and $E_{peak}$ is the peak energy in the $\nu F_{\nu}$
spectrum (when $\beta < -2$).

Since we are here caring about the high-energy power-law portion of
the GRB spectra, and also because one possible source of systematic
error in the spectral parameter determination arises in selecting
different spectral models for different bursts (Preece et al. 2002),
only the spectral parameters of those 'Band' function fitted spectra
are selected for our analysis.

One of our major concerns is to select the sample of spectra for
which $\beta$ is reliably determined. The BATSE burst
signal-to-noise ratio decreases at higher energies as a result
of fewer photon flux and the decreased detector efficiency. In
particular, $\beta$ may not be well determined if $E_{peak}$ is
close to the higher limit of the LAD energy range, $E_{max}$
($\approx$ 2 MeV) (Preece et al. 1998),
thus we must choose those spectra with $E_{peak}$ much
lower than $E_{max}$. Therefore, we select the
spectrum for which 100 keV $< E_{peak} <$ 200 keV and the error in
$\beta$ is less than 0.1 $|\beta|$. This gives a total sample of 395
spectra for 78 bursts.

\subsection{Distribution of $p$ and its narrowing}

For electrons' distribution given by a power-law:
\begin{equation} N(\gamma_e) \propto \gamma_e^{-p}, \,\,{\rm for} \,\,\,\gamma_e > \gamma_{min},
\end{equation}
the emergent high energy synchrotron spectrum is asymptotically a
power law function: $F_{\nu} \propto \nu^{-(p-1)/2}$ for $\nu_m <
\nu < \nu_c$ (``slow cooling'' regime) and $\propto \nu^{-p/2}$ for
$\nu
> \nu_c$ (``fast cooling'' regime), where $\nu_m = \nu_{syn}
(\gamma_{min})$ is the synchrotron injection frequency, and $\nu_c =
\nu_{syn} (\gamma_{c})$ is the synchrotron cooling frequency above
which the synchrotron energy loss becomes important.

The spectral index, $p$, of shock accelerated electrons is
associated with the high-energy power-law photon index, $\beta$,
of GRB photon spectrum, by either $\beta= -p/2-1$ (``fast
cooling'' regime) or $\beta= -(p+1)/2$ (``slow cooling'' regime)
depending on relative positions of $\nu_m$ and $\nu_c$ and on
which portion of the spectrum is detected. There is one regime,
$\nu_c < \nu < \nu_m$, in which $\beta = -3/2$, independent on
$p$. This case can be ruled out by discarding those spectra with
$\beta \ge -3/2$ from our sample. We found only one with $\beta
\ge -3/2$ in the BATSE sample of 395 spectra and discarded it.

Piran (2004) argues that the fast cooling
must take place during the GRB prompt phase and the reasons are:
(i) the
relativistic shocks must radiate their energy efficiently, to avoid
a serious inefficiency problem; (ii) the electrons must cool rapidly
in order that the fast variability could be observed. But there is
no firm evidence to date that could rule out the slow cooling case
for the GRB itself, since it is difficult to measure the values of
$\gamma_c$ and $\gamma_{min}$ for a specific burst. Thus in our
analysis, we assume that each GRB spectrum above $E_{peak}$ could be
in either slow cooling or fast cooling regime, so as to minimize the
width of $p$ distribution.

First we plot distribution of $p$ by assuming all spectra are in
fast cooling regime. Then we make the distribution narrower by relaxing
this constraint. Basically the narrowing process is to move some left-hand
part of the distribution to the right by adding 1 to $p$ and assuming this
part of sample are in slow cooling regime, since there is a difference of 1
about $p$ value between the two regimes. The algorithm used is described
below.

Several algorithms are implemented to get the narrowest
distribution. In the most straightforward one, each spectra has the
freedom of calculating $p$ from $\beta$ either in ``fast cooling''
or ``slow cooling'' regime, so the number of possible distributions
is $2^N$ for a sample of $N$ spectra. The distribution having the
smallest standard deviation is chosen as the narrowest one. This
algorithm works well only for $N<20$ because of the computer
program' running time. For $N>20$, we divide the overall range of
the sample's $\beta$ distribution into 20 equal-width bins and treat
the spectra with $\beta$'s located in each bin indistinguishably.
Then we apply the first algorithm to the 20 bins. In an alternative
algorithm, we start with the histogram of $p$ calculated in the
``fast cooling'' regime and mark a demarcation line within and close
to the lower limit of the range of $p$. Then all $p$'s at left to
the line in the histogram are moved to right by adding 1 to $p$, and
the new histogram's standard deviation is calculated. Repeat it after
shifting the demarcation line rightward by a step of 0.01 on the
$p$-axis. Finally the smallest standard deviation, hence the
narrowest distribution, is found. It turns out that both algorithms
give the same results for most of the samples presented in this
paper. For one sample where minor difference exists between two
algorithms' results, we use the narrower one.

We show the results of the analysis for BATSE bursts in Figure 1. 
Note that all the errors presented in this paper are at 1 $\sigma$ level.
The parent distribution of $p$ for BATSE bursts has a width of 0.54
at a 14-$\sigma$ confidence level. The method that estimates the
mean and the width of the parent distribution of $p$ is described
below. Note that the mean value of $p$ is $\approx$ 3, substantially
larger than that for the distribution before the minimization, which
is an artifact of choosing some of the spectra to be in the ``slow
cooling'' regime, equivalent to moving the left part of the
histograms in the upper panels rightward, in order to minimize the
width of the distribution.

\subsection{Statistical description of the narrowness of $p$'s distribution}

The observed distribution of $p$ plotted in Fig. 1 is a convolution
of the measurement error distribution and the true distribution (or
parent distribution) of $p$'s. What we want to know is the true
distribution $p$. We use the maximum likelihood method to estimate
the true $p$-distribution. Let us say the true distribution of $p$
is Gaussian,
\begin{equation} P(p) \ = \frac{1}{\sqrt{2\pi}\sigma_p}
\exp[-\frac{1}{2}\frac{(p-\bar{p})^2}{\sigma_p^2}]. \end{equation}
Further, we assume the measurement errors have Gaussian
distributions too. Then the probability distribution for any one
measurement $(p_i, \sigma_i)$ is the convolution of two Gaussians,
which is the Gaussian
\begin{equation} P(p_i, \sigma_i, \bar{p}, \sigma_p) = \frac{1}{\sqrt{2\pi}(\sigma_p^2+\sigma_i^2)^{1/2}}
\exp[-\frac{1}{2}\frac{(p_i-\bar{p})^2}{\sigma_p^2+\sigma_i^2}].
\end{equation}
The likelihood function for the set of $n$ measurements ${p_i,
\sigma_i}$ is
\begin{equation} L = \prod\limits_{i=1}^{n} \frac{1}{\sqrt{2\pi}(\sigma_p^2+\sigma_i^2)^{1/2}}
\exp[-\frac{1}{2}\frac{(p_i-\bar{p})^2}{\sigma_p^2+\sigma_i^2}].
\end{equation}
The principle of the Maximum Likelihood Estimate is that, the best
estimates of $\bar{p}$ and $\sigma_p^2$ are the ones that maximize
$L$. Take
\begin{equation} l = ln L = -\frac{1}{2} \sum\limits_{i}^{n} \frac{(p_i-\bar{p})^2}{\sigma_p^2+\sigma_i^2} - \frac{1}{2}
\sum\limits_{i}^{n} ln (\sigma_p^2+\sigma_i^2) , \end{equation}
then the maximum occurs when the following equations
\begin{equation} \left. \frac{\partial l}{\partial \bar{p}} \right|_{\hat{\bar{p}}, \hat{\sigma_p^2}} = 0, \end{equation}
\begin{equation} \left. \frac{\partial l}{\partial (\sigma_p^2)} \right|_{\hat{\bar{p}}, \hat{\sigma_p^2}} = 0 \end{equation}
have their solution at $\bar{p}=\hat{\bar{p}}$ and $\sigma_p^2 = \hat{\sigma_p^2}$, where `` $\hat{}$ '' symbolizes the best estimation of the parameters.
If we assume that the distribution of $\hat{\bar{p}}$ and
$\hat{\sigma_p^2}$ are both Gaussian, then one can show that the
variances of $\hat{\bar{p}}$ and $\hat{\sigma_p^2}$ are
\begin{equation} \sigma_{\hat{\bar{p}}}^2 = - \bigg[ \left.\frac{\partial^2 l}{\partial \bar{p}^2} \right|_{\hat{\bar{p}}, \hat{\sigma_p^2}} \bigg]^{-1}, \end{equation}
\begin{equation} \sigma_{\hat{\sigma_p}^2}^2 = - \Bigg[ \left.\frac{\partial^2 l}{\partial (\sigma_p^2)^2} \right|_{\hat{\bar{p}}, \hat{\sigma_p^2}} \Bigg]^{-1}, \end{equation}
respectively. So the best estimate of the parameters of true distribution of $p$
are obtained by numerically solving equations (7) and (8), and their associated
errors are calculated through equations (9) and (10).

\section{Systematic errors in $\beta$}

\subsection{The `Band' function fit to the spectra}

Preece et al. (2000) carried out a Band function fit to GRB
spectra observed by BATSE, and this way determine the high energy
power-law index ($\beta$) and the random error in $\beta$ due to
error in the observed spectral energy distribution. There is also a
systematic error in $\beta$ resulting from the finite bandwidth of
the BATSE detector, which was not reported in Preece et al., and we
estimate it here. The purpose of this exercise is to estimate the
contribution of this systematic error, and its dependence on the
peak of the spectrum ($E_{peak}$), to the dispersion in the
$p$-distribution.

The systematic error arises because the synchrotron spectrum
does not make a sharp transition from one power-law index to another
when one crosses a characteristic frequency. In particular, the
steepening of the spectrum to $\nu^{-p/2}$ above the synchrotron and
cooling frequencies does not occur suddenly at $E_{peak}$, but instead
the spectrum approaches this theoretical value asymptotically at $E
\gg E_{peak}$.

Since the spectrum is observed in a finite energy range, the
measured spectral index will always be somewhat smaller than the
true asymptotic value by an amount that depends on the ratio of
$E_{max}$ and $E_{peak}$ ($E_{max}$ is the highest energy photon
that the detector is sensitive to). The larger the
$E_{max}/E_{peak}$ is the smaller the systematic error in $\beta$
would be, and this dependence on $E_{peak}$ causes some broadening
of the observed $\beta$ distribution.

To estimate this systematic error we generate synthetic
spectra with different values for $E_{peak}$, and carry out a Band
function fit to the synthetic spectra to determine $\beta$ and its
deviation from the true asymptotic value.

The synthetic synchrotron spectra is calculated for a
relativistic homogeneous shell. The electron distribution function
behind the shock is taken to be a single power-law function: 
$N(\gamma_e) \propto \gamma_e^{-p}$, for $\gamma > \gamma_{min}$, where
$m_e c^2 \gamma_{min}$ is the minimum electron energy after they cross
the shock front. The magnetic field in the shell is taken to be
uniform and the energy density is the field is some fraction
($\epsilon_B$) of the thermal energy density of the shocked fluid;
$\gamma_{min}$ and $\epsilon_B$ are chosen so that the peak of the
spectrum, $E_{peak}$, is at some desired value. As electrons move
down-stream from the shock front they cool via the synchrotron and
inverse-Compton processes, and their distribution function is
modified. We calculate the effect of this cooling on electron
distribution functions using a self consistent scheme described in
Panaitescu \& M$\acute{e}$sz$\acute{a}$ros (2000) and McMahon et al.
(2006).

The synchrotron spectrum, for a given electron distribution,
in the shell comoving frame is calculated as described in detail by
Sari et al. (1998) (also see section 2.2). The spectrum in the
observer frame is calculated by integrating the spectral emissivity
in the comoving frame over the equal-arrival-time surface as
described in Kumar \& Panaitescu (2000).  Errors are then added to
this spectrum in a way that mimics the real GRB spectrum.

The synthetic spectrum for a known $p$ is fitted to the Band
function in a finite energy range corresponding to the BATSE energy
coverage. By varying $E_{peak}$ of the generated spectra we 
determine the discrepancy between fitted value and ``true'' 
value of $\beta$ as a function of $E_{max}/E_{peak}$. 
The results are shown in Figure 2. We find the fit
always gives a smaller $\beta$ (in absolute value) than 
the true asymptotic value and that the ``observed'' $\beta$ 
does indeed depend on $E_{max}$. The error in
$\beta$ is about 10\% when $E_{max}/E_{peak}$ is order unity,
whereas the error is $\sim$5\% when $E_{max}/E_{peak}$ $\sim$ 20.
The error also depends on the $p$ value as shown in Figure 2; for
$E_{peak}$ located between 100 keV to 200 keV, $E_{max}$= 1.8 MeV,
and $p$= 2.5, the contribution of this systematic error to the
dispersion in $\beta$ is less than 1.3\% -- the corresponding
contribution to the dispersion in $p$ is $\sigma_p < 0.03$.

We have also carried out a similar calculation for the
synchrotron self- inverse-Compton (SSC) spectrum for a population of
synchrotron electrons. The incident photons are the synchrotron
photons due to the same population of electrons that contribute to
inverse-Compton scatterings. The synchrotron radiation is taken to
be homogeneous and isotropic in the shell comoving frame, and its
spectrum is calculated as described above. The overall SSC spectrum
is obtained by the convolution of the synchrotron spectrum and
electron energy distribution using equation (7.28 a) in Rybicki \&
Lightman (1979). The curvature in the SSC spectrum is due to the
convolution of the incident spectrum and the electron distribution,
and we find that the asymptotic value for the SSC power-law index is
reached when $E_{max}/E_{peak} \sim$ 100. For this reason we find
that for the SSC case, the systematic error in $\beta$ is $\sim$
13\% for the typical $E_{max}/E_{peak}$ in BATSE bursts. The
dispersion in $p$ caused by $E_{peak}$ being distributed between
$E_{peak}$= 100 keV to $E_{peak}$= 200 keV is, however, small --
$\sigma_p < 0.04$.

These results show that the discrepancy between the fitted value and
the ``true'' value of $\beta$ is small and dependent on
$E_{max}/E_{peak}$, but its dependence on $E_{max}/E_{peak}$ is too
small to account for the observed dispersion in the $p$
distribution.

\subsection{Time-averaging effect}

Another source of systematic error in $\beta$ is the time-averaging
of multiple spectra undergoing spectral evolution, i. e., $E_{peak}$
evolving with flux (Ford et al. 1995, Crider et al. 1999). The
flux-weighted time-averaging of multiple `Band' spectra may distort
the intrinsic high-energy power law.

To examine this effect, we select BATSE time-resolved spectra with
$E_{peak}$ in 100 - 200 keV and in 200 - 300 keV, respectively,
divide them into non-evolving groups and evolving groups, and
analyze their $p$ distributions separately. The results are shown in
Table 1. We find the evolving spectra groups tend to have flatter $p$
or $\beta$, which may be an outcome the time-averaging effect. But
the widths of $p$ distributions for two groups are consistent with
each other, showing that the time-averaging does not contribute to
observed dispersion in $p$ in Figure 1.

The time-averaging effect is further examined when we use an early
BATSE spectral catalog by Band et al. (1993) in which the
time-integrated spectrum of each burst is fitted with the `Band'
function. We restrict our samples to those with $E_{peak} \le$ 300
keV, and error in $\beta$ less than 0.1 $|\beta|$, which gives a
sample of 32 spectra from the catalog of 54 GRBs. The $p$
distribution is shown in Figure 3. Comparing with Figure 1, one can see
that it has approximately the same $\sigma_p$ as that for the
time-resolved GRB spectra. This supports that the time-averaging
effect has no impact on the observed dispersion in $p$.

\section{$p$-distributions for BeppoSAX GRBs and HETE-2 XRFs, XRRs and GRBs}

We also analyzed a sample of 11 GRBs observed by BeppoSAX. The
combined (2 - 700 keV) Wide Field Cameras (WFC) and Gamma-Ray Burst
Monitor (GRBM) spectra for these bursts are fitted with the `Band'
function by Amati et al. (2002). The narrowest distribution of $p$
for this sample is shown in Figure 4 left panel. It has the same
estimated mean value of $p$ as in the BATSE bursts, and the width of
the parent distribution for $p$ is consistent with that for the
BATSE bursts. The larger errors in $<p>$ and $\sigma_p$ are due to
the smaller size of the BeppoSAX sample.

Sakamoto et al. (2005) present a catalog of X-ray flashes (XRFs),
X-ray-rich (XRR) GRBs and GRBs observed by HETE-2 WXM (2 - 25 keV)
and FREGATE (7- 400 kev) instruments. Among 45 bursts in the
catalog, 16 bursts have measured high-energy power-law photon index,
$\beta$, which is obtained through the spectral fit with the `Band'
function or a single power-law model. For those XRF spectra fitted
by a single power law, it is found that $\beta <$-2. Sakamoto et al.
(2005) explain this as that we are observing the high-energy
power-law portion of their ``Band''-function spectra. Two GRBs (GRB
020813 and 030519) for which ``Band'' model is used have $E_{peak}$
lying near or above the upper limit of FREGATE energy range, so we
exclude them here. We also exclude XRF 030528 which has a large
error in $\beta$. The final HETE-2 sample we considered comprises 7
XRFs, 4 XRRs and 2 GRBs. The $p$ distribution is shown in Figure 4
right panel.

\section{$p$-distribution for X-ray afterglows}

We also determine the distribution of $p$ during the X-ray
afterglows. We use a catalog of X-ray afterglows
observed by BeppoSAX compiled by De Pasquale et al. (2005) and a
catalog of X-ray afterglows observed by Swift (O'Brien et al.
2006). In De Pasquale et al. (2005)'s catalog, 15 X-ray afterglow
spectra are fitted with a Galactic-and-extragalactic absorbed single
power law. We use 14 out of them for our analysis and exclude GRB
000210 which has an extremely large error in measured $\beta$.
In O'Brien et al. (2006)'s Swift catalog of 40 X-ray
afterglows, we select samples with small errors, $\sigma(\beta_i) <
0.1 |\beta_i|$, and discard a sample with extremely large $|\beta|$
(= 5.5). We also discard 4 samples with $\beta_i \ge -3/2$ because
these $\beta$ values indicate the X-ray band probably lies between
$\nu_c$ and $\nu_m$ ($\nu_c < \nu_X < \nu_m$), where the asymptotic
spectral index is $\beta = -3/2$ and carries no information about
$p$. This gives 28 samples from the catalog.

The $p$-distributions for the two afterglow samples are
shown in Figure 5. For the BeppoSAX afterglows, the narrowest
distribution is consistent with a $\delta$-function distribution
within 1 $\sigma$ errors; for the Swift afterglows, it is not. The
smaller estimated width of the parent $p$-distribution for BeppoSAX
afterglows, we suspect, is due to larger errors in photon indices
$\beta_i$ of the BeppoSAX sample, $<\sigma_i(\beta)>$= 0.26, than
the Swift sample which has $<\sigma_i(\beta)>$ = 0.10.

\section{Distribution of $p$ in Blazars and pulsar wind nebulae}

\subsection{Blazars}

Blazars are active galactic nuclei with
the relativistic jet pointed toward us.
The nonthermal spectra of blazars are due to synchrotron or/and inverse Compton
emission of relativistic electrons accelerated by shocks within the jet
(Blandford \& K\"onigl 1979, Sikora et al. 1994).

Donato et al. (2005) present a spectral catalog of BeppoSAX six years
of observations of Blazars at 0.1 - 50 keV. This catalog comprises three
classes of blazars, namely low-luminosity sources
(High-energy peaked BL Lacs, or HBLs),
mid-luminosity sources (Low-energy peaked BL Lacs, or LBLs) and
high-luminosity sources (Flat Spectrum Radio Quasars, or FSRQs). The
three classes have
different locations of synchrotron peak. X-rays from
HBLs are likely to be above the peak of synchrotron spectrum, thus have
steep X-ray spectra ($\beta <$ -2), while FSRQs and LBLs in X-ray band
have more contribution from inverse Compton component and thus have flatter spectra.

From this catalog we use 44 spectra of 33 HBLs (some sources have
multi-epoch spectra) that are best fitted by single power-laws. The
errors of fitted photon indices reported in Donato et al. (2005) are
at 90\% confidence level which we convert to 1-$\sigma$ errors. The
distribution of $p$ derived from their photon spectral indices is
shown in Figure 6. We find that the distribution of $p$ for blazars is
not consistent with a $\delta$-function distribution: $\sigma_p=
0.22\pm 0.03$ after the narrowing.

\subsection{Pulsar wind nebulae}

Power-law nonthermal spectra are also often observed in pulsar wind
nebulae (PWNs) of rotation-powered pulsars. The nebular emission is
the synchrotron radiation from charged particles heated by the
termination shock in relativistic outflow (winds) from the pulsar
(see Arons (2002) for a review). Gotthelf (2003) presents a catalog
of nine bright Crab-like pulsar systems with Chandra observations
and the photon indices of pulsar nebulae, $\beta_{PWN}$, and their
90\% confidence errors are provided. We derive the distribution of
$p$ from $\beta_{PWN}$ with the $\beta_{PWN}$ errors converted into
1 $\sigma$ errors and find that  $\sigma_p= 0.59\pm 0.15$, $<p>$=
$1.72\pm0.20$ assuming the X-ray band is in the fast cooling regime.
After narrowing, the narrowest distribution has $\sigma_p= 0.24\pm
0.07$, $<p>= 2.04\pm0.09$.

\section{Summary and Discussions}

Motivated by theoretical calculations and numerical
simulations showing that the shock-accelerated electrons in relativistic
shocks have a power-law distribution with an universal index
$p \simeq 2.2-2.3$, we have determined the values of $p$ from $\gamma$-ray and
X-ray spectra  for a number of relativistic sources such as GRBs
(prompt emissions and afterglows), blazars and pulsar wind nebulae.

The maximum likelihood estimate of the width of the parent
distribution for GRB prompt emission is found to be quite broad,
$\sigma_p=0.51\pm 0.02$; the probability that the distribution is
consistent with a $\delta$-function is extremely small, and
therefore this result does not support that there is an universal
$p$.

We have considered the systematic errors in photon index due to
spectra fit and time averaging of spectra and their contributions to
the scatter in $p$ distribution. We have shown that those
contributions are very small for GRBs and can not explain the
scatter in $p$ distribution.

For X-ray afterglows of GRBs, the $p$-distribution of the
BeppoSAX sample can not rule out a possibility that the parent
distribution is a $\delta$-function distribution; however, a larger
sample of Swift afterglows is inconsistent with a
$\delta$-function parent distribution. We point out that the smaller
width of parent distribution for the BeppoSAX sample is due to its
larger measurement error in $\beta$.

Analysis of 44 blazar spectra and 9 pulsar wind nebulae shows that
the distributions of $p$ for blazars and pulsar wind nebulae (PWNe)
are also broad, not consistent with a $\delta$-function
distribution.

Possible situations in which the ``universality'' of $p$ could
break are: (i) The shock is Mildly relativistic (cf. Kirk et al.
2000); (ii) The magnetic field is oblique to the shock normal
(Baring 2005); (iii) The nature and strength of the downstream
magnetic turbulence are varying (Ostrowski \& Bednarz 2002,
Niemiec \& Ostrowski 2004). A
non-Fermi acceleration in a collisionless plasma shock was studied
by Hededal et al. (2004), in which electrons are accelerated and
decelerated instantaneously and locally, by the electric and
magnetic fields of the current channels formed through the Weibel
two-stream instability. It is not known whether an
``universality'' of $p$ could hold for this mechanism. The
``universality'' of $p$ might not happen in non-shock
accelerations; for instance, in an alternative model for GRBs
(Lyutikov \& Blandford 2003), the energy is carried outward via
magnetic field or Poynting flux. The particles accounting for the
$\gamma$-ray emissions are accelerated by magnetic field
reconnection which may also produce a power-law spectra of
accelerated particles with a variable $p$ (however, this is still
poorly understood).

\section*{Acknowledgments}
This work was supported in part by a NSF (AST-0406878) grant and a
NASA-Swift-GI grant. RFS thank Dr. Volker Bromm for his suggestion which
helps improve this work.

\newpage

\begin{figure}
\centerline{\hbox{\psfig{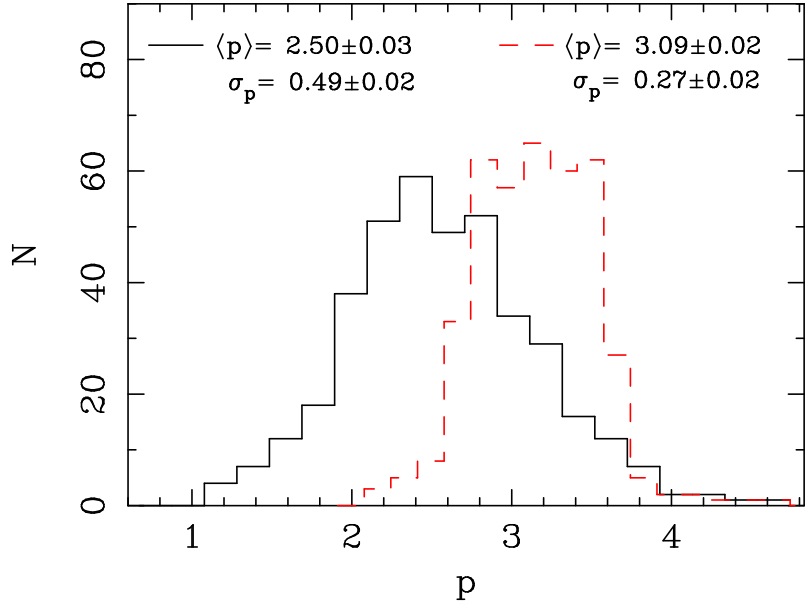}}}
\caption {Distributions of $p$ for a sample of 394 GRB spectra with 
100 $< E_{peak} <$ 200 keV. {\it Solid} line: using the relation
$p = -2\beta -2$. {\it Dashed} line: after narrowing the distribution
by using the relation of either $p = -2\beta -2$ or $p = -2\beta -1$.}
\end{figure}

\begin{figure}
\centerline{\hbox{\psfig{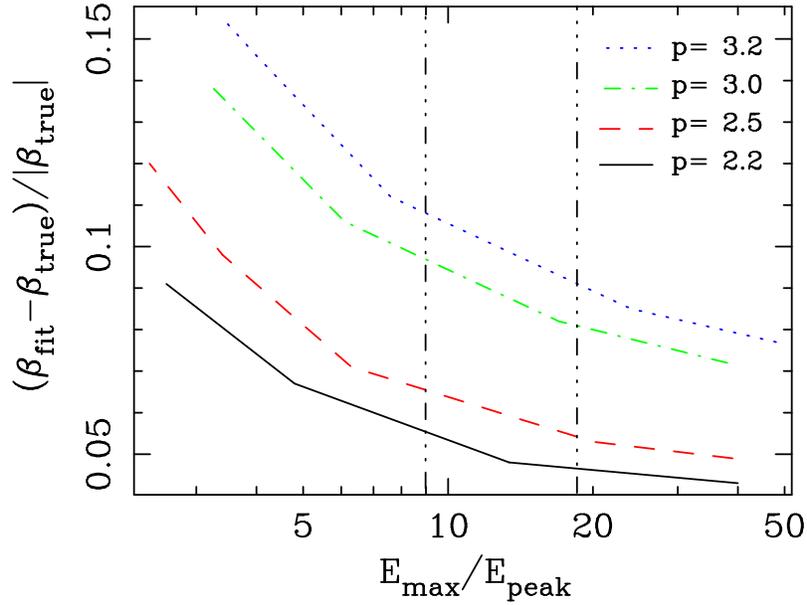}}}
\caption {The discrepancy between the fitted value and the `true'
value of $\beta$, as a function of the higher end of the fitting
energy range for the synchrotron spectra fitted by the `Band'
function. Two vertical lines mark the range of $E_{max}/E_{peak}$ corresponding
to the $E_{peak}$ range of the sample in Fig. 1.
The errors of the spectrum data are assumed to be proportional
to square root of photon counts: $\sigma(N(\nu)) \propto \sqrt{N(\nu)}$.}
\end{figure}

\begin{table}
\begin{tabular}{ccccc}
\hline
 &\multicolumn{2}{c}{$100 < E_{peak} < 200$ keV}& \multicolumn{2}{c}{$200 < E_{peak} < 300$ keV}\\
Spectra samples & Non-evolving & Evolving & Non-evolving & Evolving \\
\hline
$<p>$ & 2.86 $\pm$ 0.06 & 2.38 $\pm$ 0.03 & 2.50 $\pm$ 0.07 & 2.14 $\pm$ 0.03\\
$\sigma(p)$& 0.44 $\pm$ 0.04 & 0.47 $\pm$ 0.03 & 0.58 $\pm$ 0.06 & 0.42 $\pm$ 0.03\\
\hline
\end{tabular}
\caption{Parameters of parent distribution of $p$ for BATSE GRB
spectra samples with $E_{peak}$-evolution ($\Delta E_{peak} > 15\%
E_{peak}$) and without $E_{peak}$-evolution ($\Delta E_{peak} < 15\%
E_{peak}$), where $\Delta E_{peak}$ is the $E_{peak}$ difference
between any two {\it adjacent-in-time} spectra. All spectra are
assumed in ``fast cooling'' regime. }
\end{table}

\begin{figure}
\centerline{\hbox{\psfig{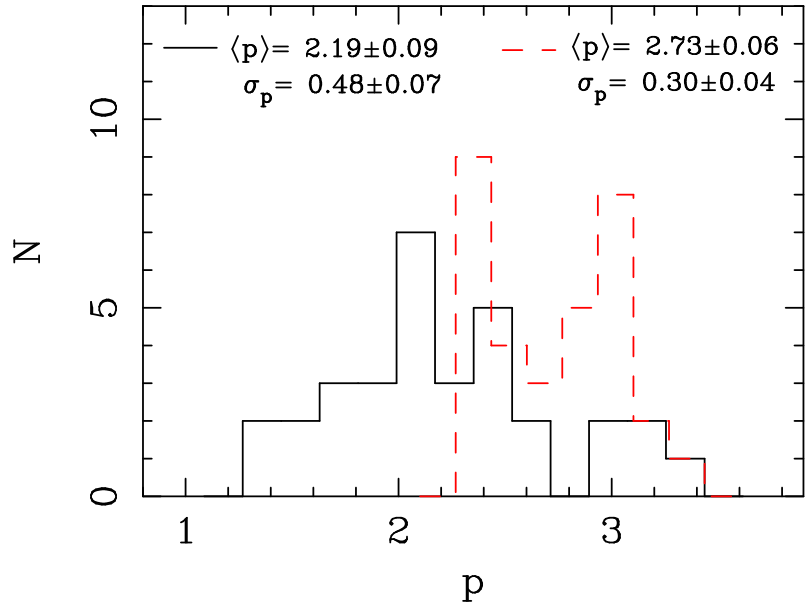}}}
\caption {The distribution of $p$ determined from 32 time-integrated GRB spectra.
{\it Solid} line: $p$ is inferred from the high-energy power-law index $\beta$
by the relation $p= -2\beta-2$. {\it Dashed} line:
the narrowest distribution of $p$ using the relation either
$p= -2\beta-2$ or $p= -2\beta-1$. $\beta$ is taken from the
`Band'-function fit by Band et al. (1993) to the time-integrated
spectrum for each burst.}
\end{figure}

\begin{figure}
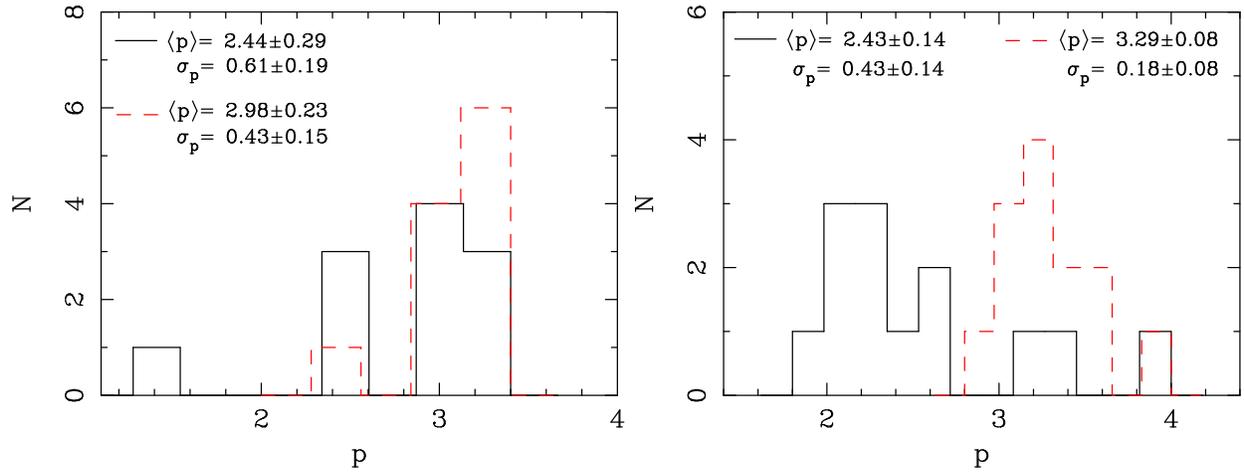

\centerline{\hbox{\psfig{file=amati02_bsax_grb.ps,width=6.2cm,angle=270}
\psfig{file=sakamoto05_hete2.ps,width=6.2cm,angle=270}}} \caption
{{\it Left}: The distributions of $p$ for 11 GRBs observed by
BeppoSAX (Amati et al. 2002); {\it Right}: The distributions of $p$
for 13 X-ray flashes, X-ray-rich GRBs and GRBs observed by HETE-2
(Sakamoto et al. 2005). {\it Solid} lines: $p$ is inferred from the
higher-energy photon index $\beta$ by the relation $p= -2\beta-2$.
{\it Dashed} lines: the narrowest distributions of $p$ using the
relation either $p= -2\beta-2$ or $p= -2\beta-1$.  }
\end{figure}

\begin{figure}
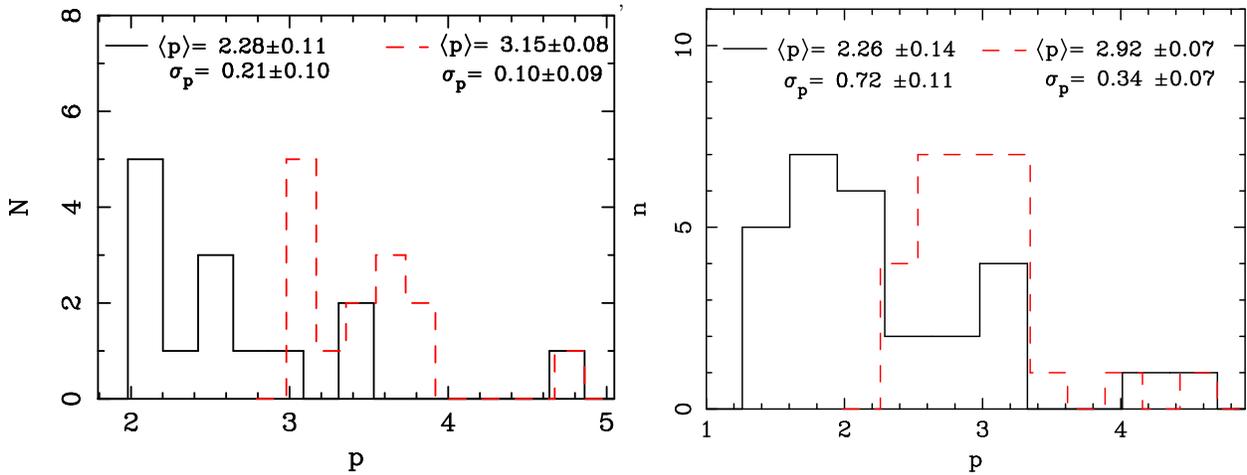

\centerline{\hbox{\psfig{file=depasquale05_bsax.ps,width=6.2cm,angle=270},
\psfig{file=obrien06_xrt.ps,width=6.2cm,angle=270}}} \caption {The
distributions of $p$ for GRB X-ray afterglows. {\it Left}: 14
afterglows are observed by BeppoSAX, taken from De Pasquale (2005).
{\it Right}: 28 afterglows are observed by Swift, taken from O'Brien
et al. (2006). {\it Solid} lines: $p$ is inferred from the photon
index $\beta$ by the relation $p= -2\beta-2$. {\it Dashed} lines:
the narrowest distribution of $p$ using the relation either $p=
-2\beta-2$ or $p= -2\beta-1$.}
\end{figure}

\begin{figure}
\centerline{\hbox{\psfig{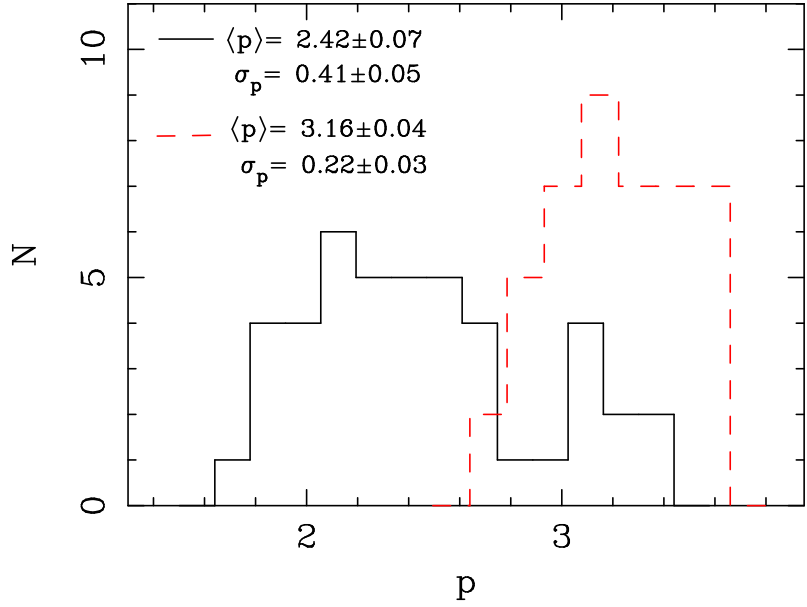}}}
\caption {The distribution of $p$ for 44 X-ray spectra of 33
blazars. {\it Solid} line: $p$ is inferred from the photon index
$\beta$ by the relation $p= -2\beta-2$. {\it Dashed} line: the
narrowest distribution of $p$ using the relation either $p=
-2\beta-2$ or $p= -2\beta-1$. $\beta$ is taken from the catalog
compiled by Donato et al. (2005).}
\end{figure}

\newpage

\end{document}